\documentclass[prb,preprint,showpacs,preprintnumbers,amsmath,amssymb,superscriptaddress]{revtex4}

\usepackage{graphicx}
\usepackage{dcolumn}
\usepackage{bm}

\begin{document}

\preprint{APS/123-QED}

\title{Complex Orbital State Stabilized by Strong Spin-Orbit Coupling in a Metallic Iridium Oxide IrO$_{2}$}

\author{Yasuyuki Hirata}
\author{Kenya Ohgushi}
\author{Jun-ichi Yamaura}
\affiliation{
Institute for Solid State Physics, University of Tokyo, Kashiwanoha 5-1-5, Kashiwa, Chiba 277-8581, Japan
}

\author{Hiroyuki Ohsumi}
\author{Soshi Takeshita}
\affiliation{
RIKEN SPring-8 Center, Kouto 1-1-1, Sayo-cho, Hyogo 679-5148, Japan
}

\author{Masaki Takata}
\author{Takahisa Arima}
\affiliation{
RIKEN SPring-8 Center, Kouto 1-1-1, Sayo-cho, Hyogo 679-5148, Japan
}
\affiliation{
Department of Advanced Materials, University of Tokyo, Kashiwanoha 5-1-5, Kashiwa, Chiba 277-8561, Japan
}


\begin{abstract}
Resonant x-ray diffraction experiments were performed for the metallic iridium oxide IrO$_{2}$. We observed anisotropic tensor of susceptibility (ATS) scattering, the spectrum of which shows a sharp contrast between the $L_{3}$ and $L_{2}$ edges. At the $L_{3}$ edge, resonance excitations were clearly observed from the core 2$p$ orbitals to both the 5$d$ $t_{2g}$ and $e_{g}$ orbitals. In contrast, the resonance mode associated with 5$d$ $t_{2g}$ orbitals was indiscernible at the $L_{2}$ edge. This contrasting behavior indicates that Ir 5$d$ $t_{2g}$ orbitals are fairly close to the $J_{\rm eff}$ = 1/2 state due to the strong spin--orbit coupling in 5$d$ transition metal ions, as in the Mott insulator Sr$_{2}$IrO$_{4}$. Our results clearly demonstrate that ATS scattering is a useful probe for investigating complex orbital states in a metallic state. Such states induce novel phenomena such as the spin Hall effect.
\end{abstract}


\maketitle

After the discovery of high-temperature superconductivity in cuprates, the correlated electronic properties have been extensively studied in on 3$d$ transition metal oxides. However, recent studies have revealed that nontrivial electronic properties emerge also in 4$d$ and 5$d$ transition metal oxides, where in addition to the crystal field and Coulomb repulsion, the strong spin-orbit coupling plays a crucial role. A good example is iridium oxides. The $t_{2g}$ orbitals are reconstructed by spin--orbit coupling into doubly degenerate $J_{\rm eff}$ = 1/2 and quadruply degenerate $J_{\rm eff}$ = 3/2 states, where the spin and orbital degrees of freedom are entangled in a complex manner.\cite{kim1} In an Ir$^{4+}$ ion with five 5$d$ electrons, the $J_{\rm eff}$ = 3/2 state is completely filled, whereas the $J_{\rm eff}$ = 1/2 state contains one electron. These $J_{\rm eff}$ = 1/2 carriers are localized in Sr$_{2}$IrO$_{4}$ with a layered-perovskite structure and CaIrO$_{3}$ with a post-perovskite structure.\cite{huang,sugahara} In these systems, the two-dimensional lattice topology makes the bandwidth narrower and consequently enhances the electron correlation effect.\cite{kim2,moon,jackeli,cairo3} This state has attracted great interest as a spin--orbit coupling induced Mott insulator.

On the other hand, iridium oxides with a three-dimensional lattice are typically good conductors.\cite{butler} A representative example is rutile-type IrO$_2$ with a crystal symmetry of $P4_{2}/mnm$. This compound has recently been reported to exhibit a large spin Hall effect, which is a novel topological transport phenomena caused by spin--orbit coupling.\cite{fujiwara} To determine the microscopic mechanism, it is essential to study the orbital state. In particular, it is important to establish whether the $J_{\rm eff}$ = 1/2 state is retained or melts in this metallic compound. Such investigations are expected to enhance our understanding of the intrinsic spin Hall effect observed in other compounds such as a platinum element. \cite{kimura}

\begin{figure}
\includegraphics[width=16cm]{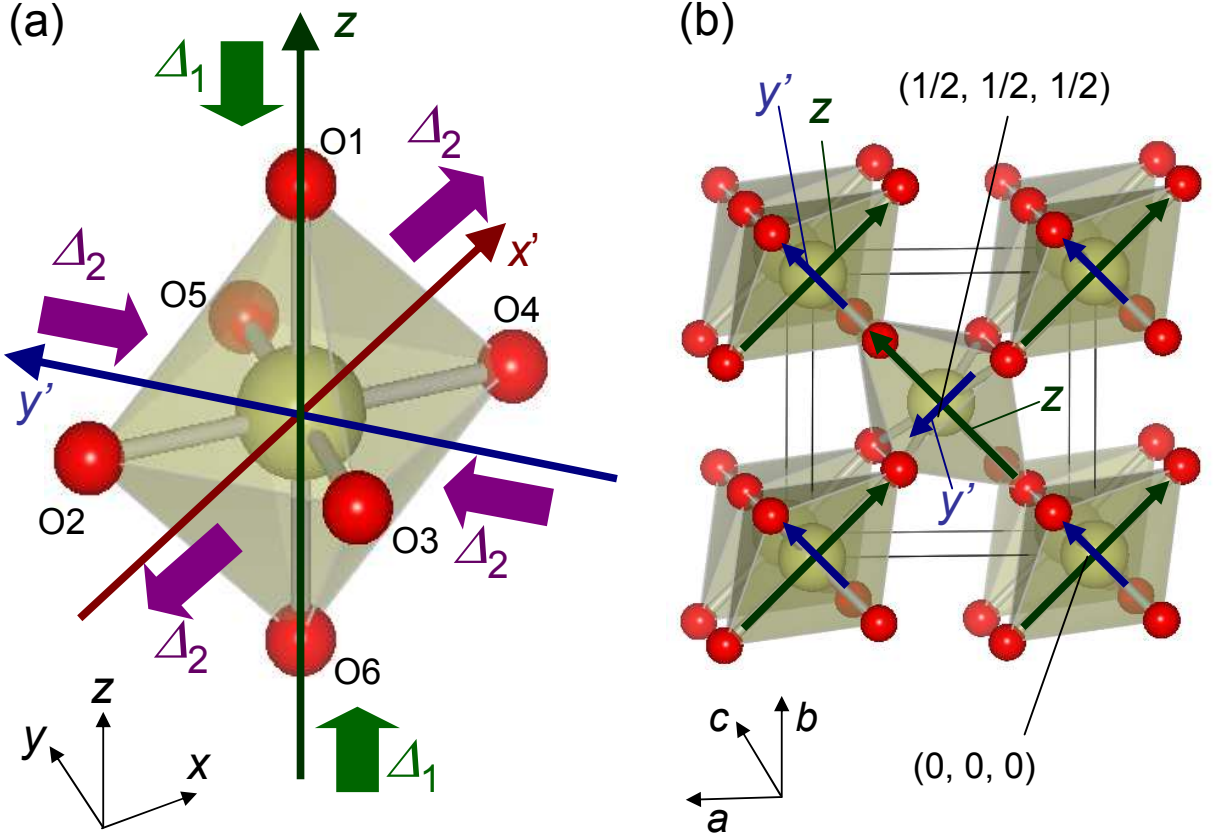}%
\caption{\label{fig:crystal} (a) Distortion of an IrO$_{6}$ octahedron in IrO$_{2}$. The compression along the $z$-axis yields the crystal field $\Delta_{1}$ and the elongation (compression) along the $x'$($y'$)-axis yields the crystal field $\Delta_{2}$. An IrO$_{6}$ octahedron is connected with the adjacent octahedra by sharing corners at O1 and O6 and by sharing the O2--O3 and O4--O5 edges. (b) Crystal structure of IrO$_{2}$. The long (short) arrows indicates the $z$- ($y'$-) axes of each IrO$_{6}$ octahedron.}
\end{figure}

Resonant x-ray scattering is a powerful technique for detecting orbital states in transition metal oxides.\cite{murakami} It is especially suitable for 5$d$ transition metal oxides because the wavelength corresponding to the dipole-allowed excitation from the core 2$p$ orbital to the valence 5$d$ orbital is comparable to lattice parameters.\cite{blume,mcmorrow} In a resonant condition, additional reflections appear at the crystallographically forbidden Bragg positions through two-types  of anomalous x-ray scattering; one is the magnetic scattering which originates from the (anti)ferromagnetic spin ordering, and the other is the anisotropic tensor of susceptibility (ATS) scattering which is associated with the antiferroic orbital structure. Among these two-types of reflections, magnetic reflections have been used to unravel the orbital states of antiferromagnets such as Sr$_{2}$IrO$_{4}$ and CaIrO$_{3}$.\cite{kim2,cairo3} However, magnetic reflections are unavailable in paramagnetic IrO$_{2}$. Instead, reflections originating from ATS scattering can be used. ATS scattering originates from the distinct local environments of Ir ions in the unit cell. In the rutile structure, IrO$_{6}$ octahedra are orthorhombically distorted (the Ir atom has a site symmetry of $m.mm$) being slightly compressed along the local $z$-axis (the Ir--O bond length is 0.196 nm along the $z$-axis and 0.200 nm in the $xy$-plane) and largely distorted in the local $xy$-plane (the O2--Ir--O3 bond angle in Fig.~\ref{fig:crystal}(a) is 75.6$^{\circ}$), giving rise to local electronic anisotropy [Fig.~\ref{fig:crystal}(a)].\cite{bolzan} The principal axis of this local anisotropy differs between the two Ir sites in the unit cell at (0, 0, 0) and (1/2, 1/2, 1/2) whose ligand octahedra are aligned to 90$^{\circ}$ rotated direction [Fig.~\ref{fig:crystal}(b)]. Consequently, ATS scattering is expected to occur at 0 0 $l$ ($l = 2n+1$).\cite{kirfel}

In this Communication, we report the nature of Ir 5$d$ orbitals in a metallic iridium oxide IrO$_{2}$ as revealed by resonant x-ray scattering measurements. We show that the intensity ratio of the ATS scattering between the Ir $L_{2}$ and $L_{3}$ absorption edges is nearly zero, which can be well explained by the selection rule expected for the $J_{\rm eff}$ = 1/2 state. Our results indicate that even in paramagnetic conducting iridium oxide IrO$_{2}$, the Ir 5$d$ orbitals are dramatically reconstructed by the strong spin--orbit coupling, as in the magnetic insulator Sr$_{2}$IrO$_{4}$.

\begin{figure}
\includegraphics[width=16cm]{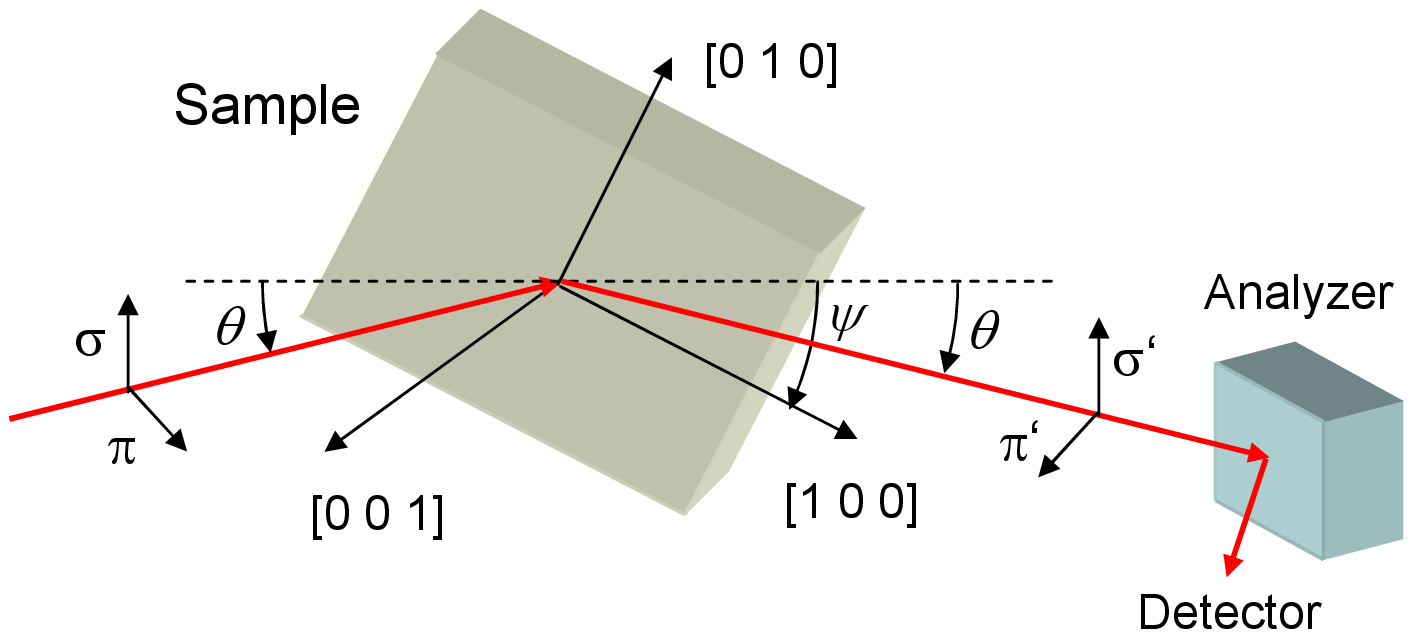}%
\caption{\label{fig:setup} Schematic of experimental setup. $\psi$ and $\theta$ denote the azimuth and incident angles, respectively. $\pi (\pi')$ and $\sigma (\sigma')$ indicate the polarization of the incident (reflected) beam.}
\end{figure}

Single crystals of IrO$_{2}$ were grown by chemical vapor transport.\cite{georg} Ir powders placed in a quartz tube were heated at 1100 $^{\circ}$C for 120 h under O$_{2}$ gas flow. Twinned IrO$_{2}$ crystals with a typical size of 5 $\times$ 2 $\times$ 1 mm$^{3}$ grew in the growth zone with a temperature of 1050 $^{\circ}$C.\cite{butler} The crystals were cut to form a single-domain crystal with an exposed (0 0 1) surface. Resonant x-ray diffraction measurements were performed at the beamline BL19LXU at SPring-8 at room temperature \cite{yabashi}. The experimental geometry is shown in Fig.~\ref{fig:setup}. The incident beam with a horizontal polarization was monochromated by a pair of Si (1 1 1) crystals to have energies of about 11.21 keV (Ir $L_{3}$ edge) and 12.82 keV (Ir $L_{2}$ edge). It was irradiated on the (0 0 1) surface of the sample with a polarization $\pi$ (within the scattering plane), which was placed such that its $a$-axis lay in the scattering plane. The azimuth angle was set to $\psi$ = 0$^{\circ}$ in this geometry ($\sigma\parallel b$). The polarization of the scattered beam was analyzed using the 0 0 8 reflection of a pyrolytic graphite crystal. We utilized a diamond phase plate to change the polarization of the incident beam from horizontal ($\pi$) to vertical ($\sigma$).\cite{hirano}

ATS scattering exhibits characteristic polarization and azimuthal angle ($\Psi$) dependences. Here, we describe these dependences according to Dmitrienko's phenomenological arguments. The matrix elements of the tensor structure amplitude matrix $\hat{F}$ for the ATS reflections at 0 0 $l$ ($l = 2n+1$) in the rutile-type structure are $F^{ab}$ = $F^{ba}$ = $F$, while the other components are zero.\cite{dmitrienko} Hence, the scattering intensity $I=|{\boldsymbol \epsilon}_{{\rm out}}^{\dagger}\hat{F}{\boldsymbol \epsilon}_{{\rm in}}|^{2}$ (where ${\boldsymbol \epsilon}_{{\rm in}}$ and ${\boldsymbol \epsilon}_{{\rm out}}$ are the polarizations of the incident and scattered x-ray beams, respectively) for each polarization condition is calculated to be $I_{\sigma\sigma'}=F^{2}\sin^{2}2\psi$, $I_{\sigma\pi'}=I_{\pi\sigma'}=F^{2}\sin^{2}2\theta\cos^{2}2\psi$, and $I_{\pi\pi'}=F^{2}\sin^{4}2\theta\sin^{2}2\psi$, where $\theta$ denotes the incident angle. $\pi (\pi')$ and $\sigma (\sigma')$ indicate the polarization of the incident (reflected) beam, as defined in Fig.~\ref{fig:setup}.

\begin{figure}
\includegraphics[width=16cm]{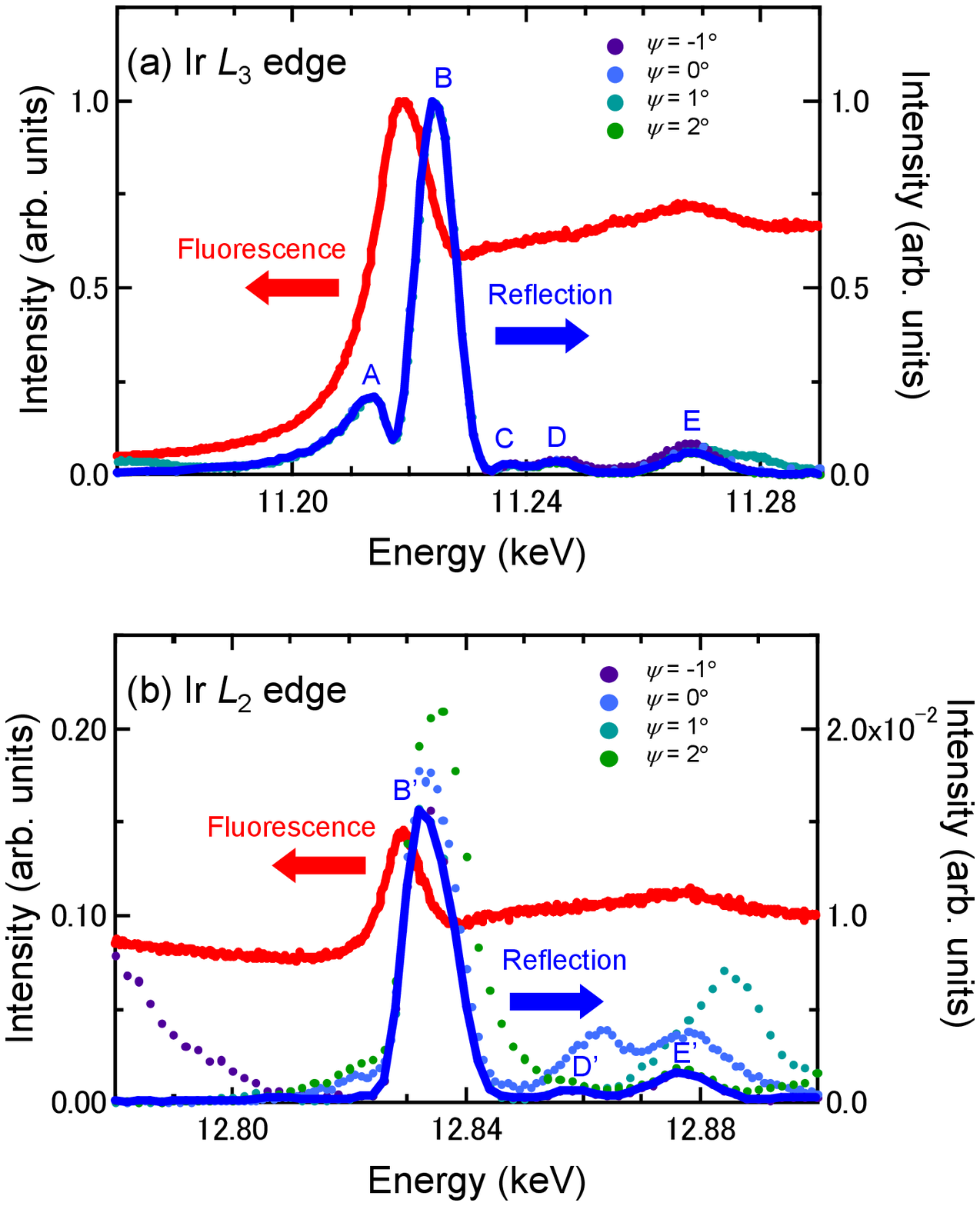}%
\caption{\label{fig:spectra}(a, b) Excitation spectra of fluorescence and 0 0 3 reflection spectra for $\pi$--$\sigma'$ channel and an azimuth angle $\psi \simeq 0^{\circ}$ for IrO$_{2}$ at the energies of the (a) $L_{3}$ and (b) $L_{2}$ edges. Dots indicate the raw data at $\psi$ = -1$^{\circ}$, 0$^{\circ}$, 1$^{\circ}$, and 2$^{\circ}$ from which the solid curves are reconstructed (see text for details). The intensities of the fluorescence and reflection spectra are normalized by the maximum values of the excitation spectrum of fluorescence and the reconstructed reflection spectrum of the $L_{3}$ edge, respectively. The peak structures labeled by A, B, C, D, E, B', D', and E' are resonance modes with excitations from the core 2$p$ orbitals to the Ir 5$d$, 6$s$, and 6$d$ bands (see text for details).}
\end{figure}

Figure \ref{fig:spectra} shows the excitation spectra of fluorescence and the energy dependence of the diffraction intensity in the $\pi$-$\sigma'$ geometry at 0 0 3, where Thomson scattering is forbidden. Reflection spectra contain several superfluous peaks originating from multiple scattering. Since multiple scattering is quite sensitive to $\psi$, its contribution could be eliminated by employing the following procedure. We measured spectra at $\psi$ = $-$1, 0, 1, and 2$^{\circ}$, chose the smallest intensity data of the four spectra for each measured energy and then reproduced the spectrum. We also measured the diffraction intensity at energies of $E$ = 11.212 and 11.224 keV by continuously changing the polarization of the incident beam from $\pi$ to $\sigma$ (data not shown) and confirmed that the polarization dependence obeys what is expected for ATS scattering.

Around the $L_{3}$ edge, which corresponds to excitation from the Ir 2$p$ $J$ = 3/2 state to 5$d$ orbitals, the corrected reflection spectrum has five-peak structure: A--E at 11.214, 11.224, 11.238, 11.245, and 11.268 keV, respectively [Fig.~\ref{fig:spectra}(a)]. Since the energy of peak A coincides with that of the absorption edge, it is assigned to be the resonant scattering associated with virtual excitation from the Ir 2$p$ $J$ = 3/2 state to the Fermi level ($i.e.$, Ir 5$d$ $t_{2g}$ orbitals). In a similar way, peak B is assigned to resonance scattering corresponding to excitation to Ir 5$d$ $e_{g}$ orbitals [Fig.~\ref{fig:spectra}(b)]. The absorption bands C, D, and E are likely related to the excitations to the Ir 6$s$ and 6$d$ bands.

We move to the $L_{2}$ edge, which corresponds to excitation from the Ir 2$p$ $J$ = 1/2 state to 5$d$ orbitals [Fig.~\ref{fig:spectra}(b)]. The edge energy is estimated to be $E$ = 12.823 keV from the excitation spectra of fluorescence. ATS scattering exhibits three peak structures: B', D', and E' at 12.833, 12.858, and 12.876 keV, respectively. Importantly, the lowest-energy peak B' is located at the inflection point of the absorption peak in the fluorescence spectrum on the higher (not lower) energy side, which is reminiscent of peak B at the $L_{3}$ edge. This indicates that peak B' is associated with the Ir 5$d$ $e_{g}$ orbitals as an intermediate state and that a $t_{2g}$ related peak is indiscernible at the $L_{2}$ edge within the experimental accuracy (less than 0.1\% of the intensity at the $L_{3}$ edge) [Fig.~\ref{fig:spectra}(b)]. The undetectable signal at the resonance condition with the $t_{2g}$ orbitals is a striking difference from the spectrum at the $L_{3}$ edge; this is conjectured to be direct evidence of the $J_{\rm eff}$ = 1/2 state in CaIrO$_{3}$. Peaks D' and E' are likely related to Ir 6$s$ and 6$d$ bands, as in the $L_{3}$ edge.

\begin{figure}
\includegraphics[width=16cm]{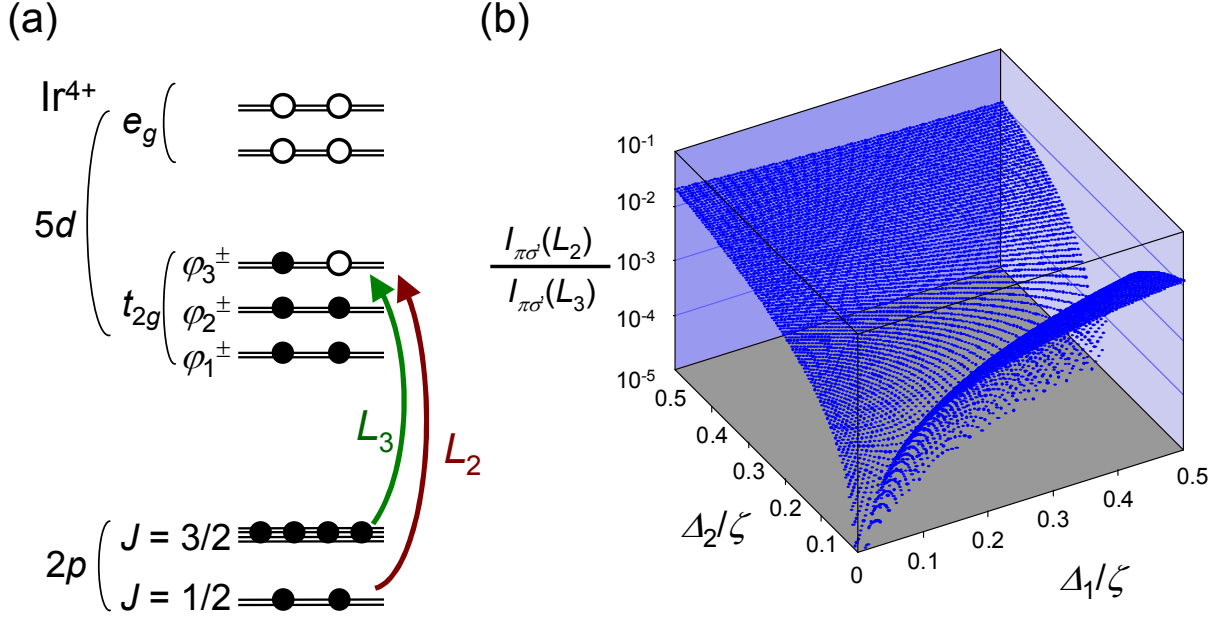}%
\caption{\label{fig:sim}(a) Schematic of resonant process between 2$p$ and 5$d$ orbitals of an Ir$^{4+}$ ion. (b) Log-scale plot of simulated scattering intensity ratio between the $L_{2}$ and $L_{3}$ edges as a function of $\Delta_{1}/\zeta$ and $\Delta_{2}/\zeta$ (see text for details).}
\end{figure}

To quantitatively evaluate the orbital state of conduction carriers in IrO$_{2}$, we performed a model calculation of the scattering amplitude. As shown in Fig.~\ref{fig:crystal}(a), IrO$_{6}$ octahedra are orthorhombically distorted. Hence, there are two crystal fields, $\Delta_{1}$ and $\Delta_{2}$: the former expresses the compression along the $z$-axis and stabilizes $xy$ orbitals; and the latter expresses the elongation along the $x'$ axis and the compression along the $y'$ axis, and stabilizes the $yz-zx$ orbital. The Hamiltonian for an Ir$^{4+}$ ion is represented by
\begin{eqnarray}
H=\left(
\begin{array}{cc}
H_{+} & 0 \\
0 & H_{-} \\
\end{array}
\right),
\end{eqnarray} 
where　 
\begin{eqnarray}
H_{\pm}=\left(
\begin{array}{ccc}
-2\Delta_{1} & \pm\frac{\zeta}{2} & -i\frac{\zeta}{2} \\
\pm\frac{\zeta}{2} & \Delta_{1} & \Delta_{2}\mp i\frac{\zeta}{2} \\
i\frac{\zeta}{2} & \Delta_{2}\pm i\frac{\zeta}{2} & \Delta_{1} \\
\end{array}
\right).
\end{eqnarray} 
Here, the basis wavefunction is ($|xy,+\rangle,|yz,-\rangle,|zx,-\rangle$, $|xy,-\rangle,|yz,+\rangle,|zx,+\rangle$) and $\zeta$ is the spin--orbit coupling constant. Each eigenstate of $H_{+}$ has its counterpart in an eigenstate of $H_{-}$, forming the Kramers doublet. Of them, two doublets with a lower energy ($\psi_{1}^{\pm}$ and $\psi_{2}^{\pm}$) are completely filled by valence electrons and the conduction band consists of $\psi_{3}^{\pm}$ orbitals [Fig.~\ref{fig:sim}(a)]. The $\psi_{3}^{\pm}$ orbitals become the $J_{\rm eff}$ = 1/2 state, $\psi_{J_{\rm eff} =1/2}^{\pm}=1/\sqrt{3}|xy,\pm\rangle\pm1/\sqrt{3}|yz,\mp\rangle+ i/\sqrt{3}|zx,\mp\rangle$ in the strong spin--orbital coupling limit ($\Delta_{1}/\zeta = \Delta_{2}/\zeta = 0$).

The atomic scattering tensor $f^{\alpha\beta}$ ($\alpha, \beta = x, y, z$) of an Ir$^{4+}$ ion is calculated using\cite{rxrs_text}
\begin{equation}
f^{\alpha\beta}\propto\sum_{n}\frac{(E_{n}-E_{0})^{3}}{\hbar\omega_{\bf k}}\frac{\langle 0|R^{\alpha}|n\rangle\langle n|R^{\beta}|0\rangle}{E_{0}-E_{n}+\hbar\omega_{\bf k}+i\Gamma_{n}/2},
\label{eq:sctensor}
\end{equation}
where $\hbar\omega_{\bf k}$ is the x-ray beam energy, $R^{i}$ is the dipole operator, and $E_{0}$ ($E_{n}$) and $|0\rangle$ ($|n\rangle$) are respectively the energy and wavefunction of the ground state (excited state). $\Gamma_{n}$ denotes the damping factor originating from the finite lifetime of the excited states. When resonant scattering associated with the $t_{2g}$ orbital at the $L_{3}$ edge occurs, one electron is virtually excited from the 2$p$ $j = 3/2$ states to one of the empty $\psi_{3}^{\pm}$ orbitals [Fig.~\ref{fig:sim}(a)]. Therefore, $f^{\alpha\beta}$ can be simplified to
\begin{equation}
f^{\alpha\beta}_{L_{3}}\propto
\sum_{j_{z}}\langle\psi_{3}^{+}|R^{\alpha}|j_{z}\rangle\langle j_{z}|R^{\beta}|\psi_{3}^{+}\rangle+\langle\psi_{3}^{-}|R^{\alpha}|j_{z}\rangle\langle j_{z}|R^{\beta}|\psi_{3}^{-}\rangle.
\end{equation}
Each dipole matrix element can be calculated by the Wigner--Eckart theorem. The matrix elements of $\hat{F}_{L_{3}}$ for the 0 0 $l$ ($l=2n+1$) reflection can then be calculated to be $F^{ab}_{L_{3}}$ = $F^{ba}_{L_{3}}$ = $f^{xx}_{L_{3}}-f^{xy}_{L_{3}}-f^{zz}_{L_{3}}$ and the other components are zero. The isotropic component of $\hat{F}_{L_{3}}$ is eliminated due to interference between the Ir$^{4+}$ ions at (0, 0, 0) and (1/2, 1/2, 1/2). These results are consistent with those of the Dmitrienko's phenomenological analysis.\cite{dmitrienko}

The intensity at the $L_{2}$ edge can be calculated in parallel with the $L_{3}$ edge. We then obtain the intensity ratio between the $L_{2}$ and $L_{3}$ edges: $I_{\pi\sigma'}(L_{2}) / I_{\pi\sigma'}(L_{3}) = \left(f^{xx}_{L_{2}}-f^{xy}_{L_{2}}-f^{zz}_{L_{2}}\right)^{2} / \left(f^{xx}_{L_{3}}-f^{xy}_{L_{3}}-f^{zz}_{L_{3}}\right)^{2}$, which are plotted as a function of $\Delta_{1}/\zeta$ and $\Delta_{2}/\zeta$ in Fig.~\ref{fig:sim}(b). This figure shows that $I_{\pi \sigma'}(L_{2})/I_{\pi \sigma'}(L_{3})$ becomes zero on a line extending from the $J_{\rm eff}$ = 1/2 state ($\Delta_{1}/\zeta=\Delta_{2}/\zeta= 0)$ to the region with large crystal-field splitting. Hence, we cannot uniquely determine the orbital character from our experimental results, $I_{\pi\sigma'}(L_{2})/I_{\pi\sigma'}(L_{3}) <$ 0.1 \%. Nevertheless, the compression along the $z$-axis in an IrO$_{6}$ octahedron is so small (the Ir--O bond length ratio is 2 \%)\cite{distortion} that we can reasonably approximate $\Delta_{1}=0$. We can then obtain the constraint $\Delta_{2}/\zeta < 0.10$, which indicates  $|\langle\psi_{J_{\rm eff} =1/2} ^{+}|\psi_{3}^{+}\rangle|^{2} > 0.997$. We conclude that the conduction electrons in IrO$_{2}$ have orbital states that is very close to the $J_{\rm eff}$ = 1/2 state.

Our results indicate the universal role of the spin-orbit coupling in iridium oxides irrespective of the conducting behavior; $J_{\rm eff}$ = 1/2 states are realized not only in insulating iridates but also in conductive iridates. This is consistent with recent x-ray absorption spectroscopy studies that revealed that Ir 5$d$ electrons in various iridium oxides including IrO$_{2}$ much prefer $J_{\rm eff}$ = 1/2 related orbitals.\cite{clancy} Such $J_{\rm eff}$ = 1/2 states can be regarded as a complex orbital  ordering, which contrasts markedly with the orbital ordering frequently observed in 3$d$ transition metal oxides such as LaMnO$_{3}$ and VO$_{2}$.\cite{murakami,vo2} In these compounds, orbital ordering is stable only in an insulating state and melting occurs if the system is metalized by, for example, heating and doping carriers. This is because the orbital ordering through the Jahn--Teller and (spin-)Peierls mechanism is driven by the energy gain of the gap opening, which inevitably makes the electronic system insulating. In contrast, the complex orbital ordering in iridium oxides is quite robust even in an itinerant system because the driving force is the spin--orbit coupling, which is inherent in Ir atoms and does not directly affect the conducting properties.

In a future study, we intend to investigate whether a metal with a complex orbital state exhibits intriguing transport phenomena. It has been theoretically argued that conducting electrons with the complex orbital state gain a non-zero Berry phase through hopping, which yields spin Hall conductance.\cite{shitade,nagaosa,sinova,kagome} This mechanism most likely explains the gigantic spin Hall effect reported in IrO$_{2}$,\cite{fujiwara}, although detailed theoretical studies are required for a deeper understanding.

Finally, we briefly comment on the crystal field splitting between the $t_{2g}$ and $e_{g}$ orbitals, 10$Dq$. 10$Dq$ is estimated to be 10 eV from the peak structure of the ATS scattering at the Ir $L_{3}$ edge. This value is somewhat larger than 3 eV in Sr$_{2}$IrO$_{4}$ deduced from resonant inelastic x-ray scattering \cite{ishii}, 7 eV in CaIrO$_{3}$ obtained by ATS scattering \cite{cairo3}, and 4 eV in IrO$_{2}$ estimated by the theoretical calculations.\cite{almeida,hamad} 10$Dq$ is unlikely to vary drastically among these compounds because the Ir--O distance in the IrO$_{6}$ octahedron is almost the same (0.196$--$0.200 nm).\cite{huang,sugahara,bolzan} To reconcile this discrepancy, more systematic and detailed studies such as resonant inelastic x-ray scattering measurements are required.

In conclusion, we have performed resonant x-ray scattering measurements of iridium oxide IrO$_{2}$. ATS scattering related to resonance with excitation from 2$p$ to 5$d$ $t_{2g}$ was clearly observed at the $L_{3}$ edge, whereas it was not observed at the $L_{2}$ edge. This contrasting behavior indicates that complex orbital states of Ir 5$d$ electrons are realized even in a metallic iridium oxide. Our results indicate that ATS scattering is a powerful technique for investigating the orbital states of metallic iridates.

The synchrotron radiation experiments were performed at BL19LXU in SPring-8 with the approval of RIKEN (Proposal No. 20110014). We are grateful to J. Matsuno, K. Fujiwara, Y. Ueda and M. Isobe for helpful discussions and experimental support. This work was supported by Special Coordination Funds for Promoting Science and Technology, Promotion of Environmental Improvement for Independence of Young Researchers and a Grant-in-Aid for Scientific Research (B) (No. 20740211).

\end{document}